\documentclass[12pt,cite,epsf,epsfig]{article}
\usepackage{epsfig}

\fontsize{12}{18}

\begin{document}
\author{S. Dev\thanks{dev5703@yahoo.com},
Sanjeev Kumar\thanks{sanjeev3kumar@gmail.com}, Surender
Verma\thanks{ s\_7verma@yahoo.co.in} and Shivani
Gupta\thanks{shiroberts\_1980@yahoo.co.in}}
\title{Phenomenology of Two-Texture Zero Neutrino Mass Matrices}
\date{Department of Physics, Himachal Pradesh University, Shimla 171005, INDIA}

\maketitle

\begin{abstract}
The generic predictions of two-texture zero neutrino mass matrices
in the flavor basis have been examined especially in relation to
the degeneracies between mass matrices within a class and
interesting constraints on the neutrino parameters have been
obtained. It is shown that the knowledge of the octant of
$\theta_{23}$, the sign of $\cos\delta$ and neutrino mass
hierarchy can be used to lift these degeneracies.
\end{abstract}

\section{Introduction}

Mass matrices provide important tools for the investigation of the
underlying symmetries and the resulting dynamics. The first step
in this direction is the reconstruction of the neutrino mass
matrix in the flavor basis. However, the reconstruction results in
a large variety of possible structures of mass matrices depending
strongly on the mass scale, mass hierarchy and the Majorana
phases. However, the relatively weak dependence on some
oscillation parameters ($\theta_{23}$ and $\delta$) results in the
degeneracy of possible neutrino mass matrices. The mass matrix for
Majorana neutrinos contains nine physical parameters including the
three mass eigenvalues, three mixing angles and the three
CP-violating phases. The two squared-mass differences ($\Delta
m^2_{12}$ and $\Delta m^2_{13}$) and the two mixing angles
($\theta_{12}$ and $\theta_{23}$) have been measured in solar,
atmospheric and reactor experiments. The third mixing angle
$\theta_{13}$ and the Dirac-type CP-violating phase $\delta$ are
expected to be measured in the forthcoming neutrino oscillation
experiments. The possible measurement of the effective Majorana
mass in neutrinoless double $\beta$ decay searches will provide an
additional constraint on the remaining three neutrino parameters
viz. the neutrino mass scale and two Majorana-type CP-violating
phases. While the neutrino mass scale will be independently
determined by the direct beta decay searches and cosmological
observations, the two Majorana phases will not be uniquely
determined from the measurement of effective Majorana mass even if
the absolute neutrino mass scale is known. Under the
circumstances, it is natural to employ other theoretical inputs
for the reconstruction of the neutrino mass matrix. The possible
form of these additional theoretical inputs are limited by the
existing neutrino data. Several proposals have been made in the
literature to restrict the form of the neutrino mass matrix and to
reduce the number of free parameters which include presence of
texture zeros \cite{1,2,3,4,5}, requirement of zero determinant
\cite{6}, the zero trace condition \cite{7} to name just a few.
However, the current neutrino oscillation data are consistent only
with a limited number of texture schemes \cite{1,2,3,4,5}. In
particular, the current neutrino oscillation data disallow all
neutrino mass matrices with three or more texture zeros in the
flavor basis. Out of the fifteen possible neutrino mass matrices
with two texture zeros, only seven are compatible with the current
neutrino oscillation data. The seven allowed two texture zero mass
matrices have been classified into three categories. The two class
A matrices of the types $A_1$ and $A_2$ give normal hierarchy (NH)
of neutrino masses. The class B matrices of types $B_1$, $B_2$,
$B_3$ and $B_4$ yield a quasi-degenerate (QD) spectrum of neutrino
masses. The single class C matrix corresponds to inverted
hierarchy (IH) of neutrino masses.

In the absence of a significant breakthrough in the theoretical
understanding of the fermion flavors, the phenomenological
approaches are bound to play a crucial role in interpreting new
experimental data on quark and lepton mixing. These approaches are
expected to provide useful hints towards unraveling the dynamics
of fermion mass generation, CP violation and identification of
possible underlying symmetries of the lepton flavors from which
realistic models of lepton mass generation and flavor mixing could
be, hopefully, constructed.

Even though the grand unification on its own does not shed any
light on the flavor problem, the Grand Unified Theories (GUTs)
provide the optimal framework in which possible solutions to the
flavor problem could be embedded. This is because the GUTs predict
definite group theoretical relations between the fermion mass
matrices. For this purpose, it is useful to find out possible
leading order forms of the neutrino mass matrix in a basis in
which the charged lepton mass matrix is diagonal. Such forms of
neutrino mass matrix provide useful hints for model building which
will eventually shed important light on the dynamics of lepton
mass generation and flavor mixing. For example, a
phenomenologically favored texture of quark mass matrix has been
presented earlier \cite{8}. In the spirit of quark-lepton
similarity, the same texture has been prescribed for the charged
lepton and Dirac neutrino mass matrices. The same texture for the
right handed neutrino mass matrix in the see-saw mechanism might
follow from universal flavor symmetry hidden in a more fundamental
theory of mass generation. Thus, the texture zeros in different
positions of the neutrino mass matrix, in particular and fermion
mass matrices, in general could be consequence of an underlying
symmetry \cite{9,10}. Such universal textures of fermion mass
matrices can, theoretically, be obtained in the context of GUTs
based on SO(10) \cite{11}. Moreover, neutrino mass matrices with
texture zeros have important implications for leptogenesis
\cite{12}.

In the present work, we examine phenomenological implications of
all possible neutrino mass matrices with two texture zeros.
Neutrino mass matrices within a class were thought to have
identical phenomenological consequences \cite{1,2,3} leading to
degeneracy of mass matrices within a class. Thus, there is a
two-fold degeneracy in class A while the neutrino mass matrices of
class B exhibit an eight-fold degeneracy since normal/inverted
hierarchies are practically indistinguishable because of
quasi-degenerate spectrum of neutrino masses in this class. We
study this degeneracy in detail and discuss possible ways to lift
this degeneracy. It is found that the deviation of atmospheric
mixing from maximality and the quadrant of the Dirac-type
CP-violating phase $\delta$ can be used to distinguish the mass
matrices within a class. We, also, note that the determination of
hierarchy will have important implications for class B neutrino
mass matrices. The prospects for the measurement of $\theta_{13}$
for class A neutrino mass matrices are quite optimistic since a
definite lower bound on $\theta_{13}$ is obtained for this class.
For neutrino mass matrices of class B, CP-violation will be near
maximal if $\theta_{13}$ is large. Definite lower bounds on the
effective Majorana mass $M_{ee}$ are obtained for neutrino mass
matrices of class B and C. Class D of neutrino mass matrices is
disallowed in our analysis.

\section{Neutrino mass matrix}

The neutrino mass matrix, $M$, can be parameterized in terms of
three neutrino mass eigenvalues ($m_{1}$, $m_{2}$, $m_{3}$), three
neutrino mixing angles ($\theta_{12}$, $\theta_{23}$,
$\theta_{13}$) and one Dirac-type CP violating phase, $\delta$. If
neutrinos are Majorana particles then there are two additional CP
violating phases $\alpha$, $\beta$ in the neutrino mixing matrix.
The complex symmetric mass matrix $M$ can be diagonalized by a
complex unitary matrix $V$:
\begin{equation}
M=VM_{\nu}^{diag}V^{T}
\end{equation}
where $M_{\nu}^{diag}=Diag \{m_1,m_2,m_3\}$. The neutrino mixing
matrix $V$ \cite{13} can be written as
\begin{equation}
V\equiv U P=\left(
\begin{array}{ccc}
c_{12}c_{13} & s_{12}c_{13} & s_{13}e^{-i\delta} \\
-s_{12}c_{23}-c_{12}s_{23}s_{13}e^{i\delta} &
c_{12}c_{23}-s_{12}s_{23}s_{13}e^{i\delta} & s_{23}c_{13} \\
s_{12}s_{23}-c_{12}c_{23}s_{13}e^{i\delta} &
-c_{12}s_{23}-s_{12}c_{23}s_{13}e^{i\delta} & c_{23}c_{13}
\end{array}
\right)\left(
\begin{array}{ccc}
1 & 0 & 0 \\ 0 & e^{i\alpha} & 0 \\ 0 & 0 & e^{i(\beta+\delta)}
\end{array}
\right),
\end{equation}
where $s_{ij}=\sin\theta_{ij}$ and $c_{ij}=\cos\theta_{ij}$. The
matrix $V$ is called the neutrino mixing matrix or PMNS matrix.
The matrix $U$ is the lepton analogue of the CKM quark mixing
matrix and $P$ contains the two Majorana phases.

The elements of the neutrino mass matrix can be calculated from
Eq. (1). Some of the elements of $M$, which can be equated to zero
in the various allowed texture zero schemes, are given by
\begin{equation}
M_{ee}=c_{13}^{2}c_{12}^{2}m_{1}+c_{13}^{2}s_{12}^{2}m_{2}e^{2i\alpha
}+s_{13}^{2}m_{3}e^{2i\beta },
\end{equation}
\begin{equation}
M_{e \mu}=c_{13}\{ s_{13}s_{23} e^{i \delta} ( e^{2 i \beta}
m_3-s^2_{12} e^{2 i \alpha} m_2 ) -c_{12}c_{23}s_{12} (m_1-e^{2 i
\alpha} m_2 ) -c^2_{12} s_{13} s_{23} e^{i \delta} m_1\},
\end{equation}
\begin{equation}
M_{e \tau}=c_{13}\{ s_{13}c_{23} e^{i \delta} ( e^{2 i \beta}
m_3-s^2_{12} e^{2 i \alpha} m_2 ) +c_{12}s_{23}s_{12} (m_1-e^{2 i
\alpha} m_2 ) -c^2_{12} s_{13} c_{23} e^{i \delta} m_1\},
\end{equation}
\begin{equation}
M_{\mu\mu}=m_1(c_{23}s_{12}+e^{i\delta}c_{12}s_{13}s_{23})^2+
e^{2i\alpha}m_2(c_{12}c_{23}-e^{i\delta}s_{12}s_{13}s_{23})^2+
e^{2i(\beta+\delta)}m_3c^2_{13}s^2_{23}
\end{equation}
and
\begin{equation}
M_{\tau\tau}=m_1(s_{23}s_{12}-e^{i\delta}c_{12}s_{13}c_{23})^2+
e^{2i\alpha}m_2(c_{12}s_{23}+e^{i\delta}s_{12}s_{13}c_{23})^2+
e^{2i(\beta+\delta)}m_3c^2_{13}c^2_{23}.
\end{equation}
It will be helpful to note from Eqs. (4-7) that the transformation
\begin{equation}
T\hspace{12pt}:\hspace{12pt}\theta_{23}\rightarrow
\frac{\pi}{2}-\theta_{23}, \delta\rightarrow\delta+\pi
\end{equation}
transforms $M_{e\mu}$ to $-M_{e\tau}$ and $M_{\mu\mu}$ to
$M_{\tau\tau}$. Therefore, if $M_{e\mu}$ vanishes for
$\theta_{23}$ and $\delta$, then $M_{e\tau}$ vanishes for
$\frac{\pi}{2}-\theta_{23}$ and $\delta+\pi$. Similarly, if
$M_{\mu\mu}$ vanishes for $\theta_{23}$ and $\delta$, then
$M_{\tau\tau}$ vanishes for $\frac{\pi}{2}-\theta_{23}$ and
$\delta+\pi$. This fact can be, gainfully, used for distinguishing
the subcategories of neutrino mass matrices within a class.

\section{Two texture zeros}

The neutrino mass matrix, $M$, in the charged lepton basis is
given by Eq. (1). Out of the fifteen possible patterns for two
texture zeros in $M$, the seven possibilities \cite{1,2}, which
are found to be consistent with the current neutrino oscillation
data, are listed in Table 1. In addition, we have, also, included
the two mass matrices of class D in the list. These mass matrices
were disallowed in some of the earlier analyses \cite{1,2}.
However, it was found in a subsequent detailed numerical analysis
by Guo and Xing \cite{3} that the neutrino mass matrices of class
D are, marginally, allowed. The expressions for the vanishing
elements in different two-zero texture schemes have been listed in
Eqs. (4-7).

\begin{table}[tb]
\begin{center}
\begin{tabular}{|c|c|}
\hline
 Type  &        Constraining Eqs.         \\
 \hline
 $A_1$ &     $M_{ee}=0$, $M_{e\mu}=0$     \\
 $A_2$ &    $M_{ee}=0$, $M_{e\tau}=0$     \\
 $B_1$ &  $M_{e\tau}=0$, $M_{\mu\mu}=0$   \\
 $B_2$ &  $M_{e\mu}=0$, $M_{\tau\tau}=0$  \\
 $B_3$ &   $M_{e\mu}=0$, $M_{\mu\mu}=0$   \\
 $B_4$ & $M_{e\tau}=0$, $M_{\tau\tau}=0$  \\
 $C$   & $M_{\mu\mu}=0$, $M_{\tau\tau}=0$  \\
 $D_1$ & $M_{\mu\mu}=0$, $M_{\mu\tau}=0$   \\
 $D_2$ & $M_{\tau\tau}=0$, $M_{\mu\tau}=0$ \\
 \hline
\end{tabular}
\end{center}
\caption{Allowed two texture zero mass matrices.}
\end{table}

The transformation T defined in Eq. (8) transforms neutrino mass
matrices of type $A_1$ to $A_2$ and neutrino mass matrices of type
$B_1$ ($B_4$) to $B_2$ ($B_3$). Therefore, the predictions for
neutrino mass matrices of types $A_1$ and $A_2$ will be identical
for all neutrino parameters except $\theta_{23}$ and/or $\delta$.
Similarly, the predictions of neutrino mass matrices for all the
four types of mass matrices in class B will be identical for all
neutrino parameters except $\theta_{23}$ and/or $\delta$. However,
the different types of mass matrices will have different
predictions for the octant of $\theta_{23}$ and the sign of
$\cos\delta$.

The two texture zeros in the neutrino mass matrix give two complex
equations viz.
\begin{equation}
M_{ab}=0,M_{pq}=0
\end{equation}
where $a$, $b$, $p$ and $q$ can take values $e$, $\mu $ and
$\tau$. Eqs. (9) can, also, be written as
\begin{equation}
m_{1}U_{a1}U_{b1}+m_{2}U_{a2}U_{b2}e^{2i\alpha
}+m_{3}U_{a3}U_{b3}e^{2i(\beta +\delta )}=0
\end{equation}
and
\begin{equation}
m_{1}U_{p1}U_{q1}+m_{2}U_{p2}U_{q2}e^{2i\alpha
}+m_{3}U_{p3}U_{q3}e^{2i(\beta +\delta )}=0
\end{equation}
where $U$ has been defined in Eq. (2). These two complex equations
involve nine physical parameters $m_{1}$, $m_{2}$ , $m_{3}$,
$\theta _{12}$, $\theta _{23}$, $\theta _{13}$ and CP-violating
phases $\alpha $, $\beta $, and $\delta $. The two mixing angles
$(\theta _{12}$, $\theta _{23})$ and two mass-squared differences
$(\Delta m_{12}^{2}$, $\Delta m_{23}^{2})$ are known from the
solar, atmospheric and reactor neutrino experiments. The masses
$m_{2}$ and $m_{3}$ can be calculated from the mass-squared
differences $\Delta m_{12}^{2}$ and $\Delta m_{23}^{2}$ using the
relations
\begin{equation}
m_{2}=\sqrt{m_{1}^{2}+\Delta m_{12}^{2}}
\end{equation}
and
\begin{equation}
m_{3}=\sqrt{m_{2}^{2}+\Delta m_{23}^{2}}.
\end{equation}
Thus, we have two complex relations relating five unknown
parameters viz. $m_1$, $\theta_{13}$, $\alpha$, $\beta$ and
$\delta$. Therefore, if one out of these five parameters is
assumed, other four parameters can be predicted. Thus, neutrino
mass matrices with two texture zeros in the flavor basis have
strong predictive power.

Solving Eqs. (10) and (11) simultaneously, we obtain
\begin{equation}
\frac{m_{1}}{m_{3}}e^{-2i\beta }=\frac{%
U_{p3}U_{q3}U_{a2}U_{b2}-U_{a3}U_{b3}U_{p2}U_{q2}}{
U_{a1}U_{b1}U_{p2}U_{q2}-U_{a2}U_{b2}U_{p1}U_{q1}} e^{2i\delta }
\end{equation}
and
\begin{equation}
\frac{m_{1}}{m_{2}}e^{-2i\alpha }=\frac{
U_{p2}U_{q2}U_{a3}U_{b3}-U_{a2}U_{b2}U_{p3}U_{q3}}{
U_{a1}U_{b1}U_{p3}U_{q3}-U_{a3}U_{b3}U_{p1}U_{q1}}.
\end{equation}
Using Eqs. (14) and (15), the two mass ratios
$\left(\frac{m_1}{m_2},\frac{m_1}{m_3}\right)$ and Majorana phases
($\alpha $, $\beta $) can be written as
\begin{equation}
\left(\frac{m_1}{m_2}\right)=\left| \frac{
U_{p2}U_{q2}U_{a3}U_{b3}-U_{a2}U_{b2}U_{p3}U_{q3}}{
U_{a1}U_{b1}U_{p3}U_{q3}-U_{a3}U_{b3}U_{p1}U_{q1}}\right| ,
\end{equation}
\begin{equation}
\left(\frac{m_1}{m_3}\right)=\left| \frac{
U_{p3}U_{q3}U_{a2}U_{b2}-U_{a3}U_{b3}U_{p2}U_{q2}}{
U_{a1}U_{b1}U_{p2}U_{q2}-U_{a2}U_{b2}U_{p1}U_{q1}}\right| ,
\end{equation}
\begin{equation}
\alpha =-\frac{1}{2}arg\left( \frac{%
U_{p2}U_{q2}U_{a3}U_{b3}-U_{a2}U_{b2}U_{p3}U_{q3}}{
U_{a1}U_{b1}U_{p3}U_{q3}-U_{a3}U_{b3}U_{p1}U_{q1}}\right)
\end{equation}
and
\begin{equation}
\beta =-\frac{1}{2}arg\left( \frac{
U_{p3}U_{q3}U_{a2}U_{b2}-U_{a3}U_{b3}U_{p2}U_{q2}}{
U_{a1}U_{b1}U_{p2}U_{q2}-U_{a2}U_{b2}U_{p1}U_{q1}}\right)-\delta.
\end{equation}
Eqs. (18) and (19) give the Majorana phases $\alpha$ and $\beta$
in terms of $\theta_{13}$ and $\delta$ since $\theta_{12}$ and
$\theta_{23}$ are known experimentally. Similarly, Eqs. (16) and
(17) give the mass ratios
$\left(\frac{m_1}{m_2},\frac{m_1}{m_3}\right)$ as functions of
$\theta_{13}$ and $\delta$. Since, $\Delta m^2_{12}$ and $\Delta
m^2_{23}$ are known experimentally, the values of mass ratios
$\left(\frac{m_1}{m_2},\frac{m_1}{m_3}\right)$ from Eqs. (16) and
(17) can be used to calculate $m_1$. This can be done by inverting
Eqs. (12) and (13) to obtain the two values of $m_1$, viz.
\begin{equation}
m_{1}=\left(\frac{m_1}{m_2}\right) \sqrt{\frac{ \Delta
m_{12}^{2}}{1-\left(\frac{m_1}{m_2}\right) ^{2}}}
\end{equation}
and
\begin{equation}
m_{1}=\left(\frac{m_1}{m_3}\right) \sqrt{\frac{\Delta m_{12}^{2}+
\Delta m_{23}^{2}}{ 1-\left(\frac{m_1}{m_3}\right)^{2}}},
\end{equation}
The two values of $m_{1}$ obtained above from the mass ratios
$\left(\frac{m_1}{m_2}\right)$ and $\left(\frac{m_1}{m_3}\right)$,
respectively, must be equal. Thus, we can constrain
($\theta_{13}$, $\delta$) plane using the experimental inputs for
$\Delta m^2_{12}$, $\Delta m^2_{23}$, $\theta_{12}$ and
$\theta_{23}$. The Dirac-type CP violating phase, $\delta $, is
given full variation from $0^{0}$ to $360^{0}$ and $\theta _{13}$
is varied from zero to its upper bound from CHOOZ experiment. We
use the current best fit values of the oscillation parameters with
1, 2 and 3 $\sigma$ \cite{14} errors given below:
\begin{eqnarray}
\Delta m_{12}^{2}
&=&7.9_{~-0.3,~0.6,~0.8}^{~+0.3,~0.6,~1.0}~\times 10^{-5}~eV^{2}~,
\nonumber \\ \Delta m_{23}^{2} &=&\pm
2.6_{~-0.2,~0.4,~0.6}^{~+0.2,~0.4,~0.6}~\times 10^{-3}~eV^{2}~,
\nonumber
\\ s_{12}^{2} &=&0.30_{~-0.02,~0.04,~0.06}^{~+0.02,~0.06,~0.10}~,
\\ s_{23}^{2}
&=&0.50_{~-0.05,~0.12,~0.16}^{~+0.06,~0.13,~0.18}~,  \nonumber \\
s_{13}^{2} &<&~0.012,~0.025,~0.040~.  \nonumber
\end{eqnarray}
Here, $'+'$ ($'-'$) sign with the value of $\Delta m^2_{23}$ is
for normal (inverted) hierarchy. The analysis \cite{14}
incorporates not only the latest long baseline data for $\Delta
m^2_{13}$ from the MINOS collaboration \cite{15} but also the
updated KamLAND and SNO data \cite{16}. We vary the oscillation
parameters within their 1, 2 and 3 $\sigma$ C.L. ranges with
uniform distributions and calculate their predictions at various
confidence levels. However, we report the correlations plots at 1
$\sigma$ C.L.. We, also, calculate the Jarlskog rephasing
invariant quantity \cite{18}
\begin{equation}
J=s_{12}s_{23}s_{13}c_{12}c_{23}c_{13}^2 \sin \delta
\end{equation}
for the allowed parameter space.

In our numerical analysis outlined above, the two squared-mass
differences ($\Delta m^2_{12}$ and $\Delta m^2_{23}$) enter as two
independent experimental inputs while the earlier analyses
\cite{1,2,3} require the ratio of two known mass-squared
differences
\begin{equation}
R_{\nu}\equiv \frac{\Delta m^2_{12}}{\Delta m^2_{23}}
=\frac{1-(\frac{m_1}{m_2})^2}{\left|1-\frac{(\frac{m_1}{m_2})^2}
{(\frac{m_1}{m_3})^2}\right|}
\end{equation}
to lie in the experimentally allowed range in order to constrain
the other neutrino parameters. Our analysis makes direct use of
the two mass-squared differences and has more constraining power
whereas the earlier analyses which make use of the ratio $R_{\nu}$
lose some of the constraining power because this procedure does
not use the full experimental information currently available with
us in the shape of two mass-squared differences. Moreover, since
$R_{\nu}$ is a function of mass ratios, it does not depend upon
the absolute neutrino mass scale. It is obvious that the two mass
ratios $\left(\frac{m_1}{m_2}\right)$ and
$\left(\frac{m_1}{m_3}\right)$ may yield the experimentally
allowed values of $R_{\nu}$ [Eq. (24)] for mutually inconsistent
values of $m_1$ [Eqs. (20) and (21)]. Such mass ratios were
allowed in the earlier analyses based upon $R_{\nu}$ \cite{1,2,3}.
However, our analysis selects only those mass ratios for which the
values of $m_1$ obtained from Eqs. (20) and (21) are identical and
the ratio $R_{\nu}$ for these values of the mass ratios is
automatically restricted to lie in the allowed experimental range.
Moreover, the definition of $R_{\nu}$
\begin{equation}
R_{\nu}\equiv \frac{\Delta m^2_{12}}{\Delta m^2_{23}}
=\frac{|m_2^2-m_1^2|}{|m_3^2-m_2^2|}
\end{equation}
used in many earlier analyses does not make use of the knowledge
of solar mass hierarchy to constrain the neutrino parameter space.
Our analysis disallows the class D of two-texture zero neutrino
mass matrices which has been marginally allowed by Guo and Xing
\cite{3}.

\section{Results and discussion}

In this section, we shall present the results of our numerical
analysis for neutrino mass matrices of classes A, B and C based
upon the approach described in the previous section. However, it
will be helpful to know the Taylor series expansion of Eqs.
(16-19) in the powers of $s_{13}$ to appreciate the numerical
results. The zeroth order terms of the Taylor series expansion
have been tabulated in Table 2.

\begin{table}[tb]
\begin{center}
\begin{tabular}{||c||c|c|c|c||}
 \hline
 Type  &       $\frac{m_1}{m_2}$        &                     $\frac{m_1}{m_3}$                     &      $\alpha$      &                    $\beta$                    \\
 \hline\hline
 $A_1$ &      $\tan^2 \theta_{12}$      &      $\tan\theta_{12}\tan\theta_{23}\sin\theta_{13}$      & $n+\frac{1}{2}\pi$ &         $\beta+\frac{\delta}{2}=n\pi$         \\
 \hline
 $A_2$ &      $\tan^2 \theta_{12}$      & $\frac{\tan\theta_{12}\sin\theta_{13}}{\tan\theta_{23}}$  & $n+\frac{1}{2}\pi$ & $\beta+\frac{\delta}{2}=(n+\frac{1}{2})\pi$ \\
 \hline
 $B_1$ &               1                &                    $\tan^2\theta_{23}$                    &       $n\pi$       &      $\beta+\delta=(n+\frac{1}{2})\pi$      \\
 \hline
 $B_2$ &               1                &               $\frac{1}{\tan^2\theta_{23}}$               &       $n\pi$       &      $\beta+\delta=(n+\frac{1}{2})\pi$      \\
 \hline
 $B_3$ &               1                &                    $\tan^2\theta_{23}$                    &       $n\pi$       &      $\beta+\delta=(n+\frac{1}{2})\pi$      \\
 \hline
 $B_4$ &               1                &               $\frac{1}{\tan^2\theta_{23}}$               &       $n\pi$       &      $\beta+\delta=(n+\frac{1}{2})\pi$      \\
 \hline
  $C$  & $\frac{1}{\tan^2 \theta_{12}}$ & $\frac{1}{\tan\theta_{12}\tan2\theta_{23}\sin\theta_{13}}$ & $n+\frac{1}{2}\pi$ &                -                              \\
  \hline
 $D_1$ & $\frac{1}{\tan^2 \theta_{12}}$ &  $\frac{\tan\theta_{23}}{\tan\theta_{12}\sin\theta{13}}$  &       $n\pi$       & $\beta+\frac{\delta}{2}=(n+\frac{1}{2})\pi$ \\
 \hline
 $D_2$ & $\frac{1}{\tan^2 \theta_{12}}$ &  $\frac{1}{\tan\theta_{23}\tan\theta_{12}\sin\theta{13}}$    & $n+\frac{1}{2}\pi$ &         $\beta+\frac{\delta}{2}=n\pi$         \\
 \hline
\end{tabular}
\caption{Zeroth order approximations for neutrino mass matrices
with two texture zeros.}
\end{center}
\end{table}

\subsection{Class A}
It can be seen from Table 2 that both the mass ratios
$\frac{m_1}{m_2},\frac{m_1}{m_3}\ll 1$ for neutrino mass matrices
of class A. So, neutrino mass matrices of class A give
hierarchical spectrum of neutrino masses. This case has, already,
been analyzed analytically \cite{5} and, here, we shall discuss
the earlier results only for the sake of completeness.

The results for $m_1$, $\alpha$, $\beta$, $\theta_{13}$ and $J$
for class A mass matrices have been summarized in Table 3 at
various confidence levels. These quantities are the same for mass
matrices of types $A_1$ and $A_2$. It can be seen from Table 3
that the 3 $\sigma$ lower bound on $\theta_{13}$ is $3.3^0$. The
range for the Majorana-type CP-violating phase $\beta$ at 1
$\sigma$ C.L. is found to be $-80^0$ - $80^0$. However, if the
neutrino oscillation parameters are allowed to vary beyond their
present 1.2 $\sigma$ C.L. ranges, the full range for $\beta$
($-90^0$ - $90^0$) is allowed. At one standard deviation, the
allowed range of $\delta$ is ($-150^0$ - $150^0$) for type $A_1$
and ($30^0$ - $330^0$) for type $A_2$ and the allowed range of
$\theta_{23}$ is ($41.8^0$ - $48.2^0$) for type $A_1$ and $A_2$
[Table 4]. Just like $\beta$, no constraint on $\delta$ is
obtained above 1.2 $\sigma$ C.L. The accuracies of the oscillation
parameters required to distinguish between the theoretical
predictions of neutrino mass matrices of types $A_1$ and $A_2$
depend on the upper bound on $\theta_{13}$. With the current
precision of the oscillation parameters, the neutrino mass
matrices of types $A_1$ and $A_2$ can be distinguished at 1
$\sigma$, 2 $\sigma$ and 3 $\sigma$ C.L. for
$\theta_{13}<5.7^0,5.2^0$ and $4.8^0$ respectively.

\begin{table}[tb]
\begin{center}
\begin{tabular}{|c|ccccc|}
\hline
 C.L.& $m_1(10^{-3}eV)$& $\alpha(\deg)$ & $\beta(\deg)$ & $\theta_{13}(\deg)$&  $J$  \\
\hline
 $1$ &$2.6$ - $4.5$  & $83$ - $97$  & $-80$ - $80 $ & $5.2$ - $6.3$ & $-0.025$ - $0.025$ \\
 $2$ &$2.1$ - $8.9$ & $77$ - $103$ & $-90$ - $90 $ & $4.2$ - $9.1$ & $-0.037$ - $0.037$ \\
 $3$ &$1.7$ - $13.4$ & $73$ - $107$ & $-90$ - $90 $ & $3.3$ - $11.5$& $-0.046$ - $0.046$ \\
\hline
\end{tabular}
\end{center}
\caption{The predictions for neutrino mass matrices of class A.}
\end{table}

\begin{table}[tb]
\begin{center}
\begin{tabular}{|c|cc|}
\hline

 1 $\sigma$ predictions      &   $\delta$          & $\theta_{23}    $  \\
\hline
 $A_1$ &$-150^0$ - $150^0$ & $41.8^0$ - $48.2^0$  \\
 $A_2$ &$30^0$ - $330^0$   & $41.8^0$ - $48.2^0$  \\
\hline
\end{tabular}
\end{center}
\caption{The predictions for $\delta$ and $\theta_{23}$ for
neutrino mass matrices of types $A_1$ and $A_2$.}
\end{table}

Fig. 1 (identical for $A_1$ and $A_2$) depicts the correlation
plots of $m_1$, $\alpha$, $\beta$ and $J$ at one standard
deviation. Fig. 1(a) shows $\alpha$ as a function of $m_1$ and
Fig1(b) shows $\beta$ as a function of $\alpha$. The large spread
in the ($m_1$,$\alpha$) plot is due to the errors in the neutrino
oscillation parameters. However, there is a very strong
correlation between $\alpha$ and $\beta$ as indicated in Table 2.
Such a correlation between the Majorana phases was noted earlier
\cite{3}. However, full ranges ($-90^0$ - $90^0$) are allowed for
the two Majorana phases in that analysis \cite{3}. In contrast, we
obtain a very narrow range for the Majorana phase $\alpha$ around
$90^0$ [Cf. Table 3]. Similarly, the Majorana phase $\beta$ is,
also, constrained if the oscillation parameters are limited to
their 1 $\sigma$ ranges. Fig. 1(c) depicts the correlation between
$J$ and $m_1$.

In Fig. 2, we depict the correlation plots of $\alpha$ and $\beta$
with $\delta$ for matrices of types $A_1$ (left panel) and $A_2$
(right panel). We, also, show the correlation plots of $\delta$
and $\theta_{23}$ with one another as well as with $\theta_{13}$
in Fig. 2. The Dirac-type CP-violating phase $\delta$ is,
strongly, correlated with the Majorana-type CP-violating phase
$\beta$ [Cf. Table 2]. It can be seen from ($\alpha$,$\delta$)
correlation plots [Fig. 2(a) and 2(b)] and ($\beta$,$\delta$)
correlation plots [Fig. 2(c) and 2(d)] that there are small
deviations in the values of $\alpha$ around $90^0$ and the
correlation between $\beta$ and $\delta$ is, almost, linear. The
fact that $\delta+2\beta\simeq 0^0$ for $A_1$ type mass matrices
and $\delta+2\beta\simeq 180^0$ for $A_2$ type mass matrices [cf.
Table 2] is apparent from the ($\beta$,$\delta$) correlation plots
[Fig. 2(c) and 2(d)]. The ($\beta$,$\delta$) correlation was noted
earlier by Xing \cite{2} in its approximate form in a different
parameterization. The ($\delta$, $\theta_{23}$) plots, clearly,
illustrate the point that the neutrino mass matrices of types
$A_1$ and $A_2$ prefer different regions on the ($\delta$,
$\theta_{23}$) plane. This is contrary to the analysis done by
Xing \cite{3} where no constraints on $\delta$ and $\theta_{23}$
have been obtained. In fact, this feature is crucial for
distinguishing mass matrices of types $A_1$ and $A_2$ which were
found to be degenerate in the earlier analyses \cite{1,2,3}. The
constraints on $\delta$ and $\theta_{23}$ are very sensitive to
the values of $\theta_{13}$. For the values of $\theta_{13}$
smaller than $1 \sigma$ CHOOZ bound, the constraints on $\delta$
and $\theta_{23}$ become stronger which can be seen from the
($\theta_{13}$, $\theta_{23}$) and ($\theta_{13}$,$\delta$)
correlation plots [Figures 2(g-j)]. For example, if
$\theta_{13}>5.6^0$, then $\theta_{23}>46^0$ (above maximal) for
type $A_1$ and $\theta_{23}<44^0$ (below maximal) for type $A_2$.
It can, also, be seen from the ($\theta_{13}$, $\theta_{23}$)
correlation plots in Fig. 2 that the deviation of $\theta_{23}$
from maximality is larger for smaller values of $\theta_{13}$.
Therefore, if future experiments measure $\theta_{13}$ below its
present $1\sigma$ bound, the neutrino mass matrices of types $A_1$
and $A_2$ will have different predictions for $\delta$ and
$\theta_{23}$ with no overlap. However, it would be difficult to
differentiate between matrices of types $A_1$ and $A_2$ if
$\theta_{13}$ is found to be above its present $1\sigma$ range. As
noted earlier, different quadrants for $\delta$ are selected for
neutrino mass matrices of types $A_1$ and $A_2$. It can, also, be
seen in Fig. 2 that the points $\delta=0,\pi$ are ruled out
implying that neutrino mass matrices of class A are necessarily
CP-violating. This feature is, also, apparent in Fig. 2 which
depict $J$ as a function of $\delta$ for types $A_1$ [Fig. 2(k)]
and $A_2$ [Fig. 2(l)].

\subsection{Class B}

It can be seen from Table 2 that the mass ratios $\frac{m_1}{m_2}$
and $\frac{m_1}{m_3}$ are approximately equal to one for neutrino
mass matrices of class B. So, neutrino mass matrices of class B
give quasi-degenerate spectrum of neutrino masses. For neutrino
mass matrices of class $B$, the ratio $\frac{m_1}{m_2}$, upto
first order in $s_{13}$, is given by
\begin{eqnarray}
B_1\hspace{12pt}:\hspace{22pt}\frac{m_1}{m_2}=1+\frac{c_{23}}
{c_{12}s_{12}s^3_{23}}s_{13}\cos\delta,  \nonumber\\
B_2\hspace{12pt}:\hspace{22pt}\frac{m_1}{m_2}=1-\frac{s_{23}}
{c_{12}s_{12}c^3_{23}}s_{13}\cos\delta,  \nonumber\\
B_3\hspace{12pt}:\hspace{11pt}\frac{m_1}{m_2}=1-\frac{c_{23}\tan^2\theta_{23}}
{c_{12}s_{12}s^3_{23}}s_{13}\cos\delta,  \nonumber\\
B_4\hspace{12pt}:\hspace{13pt}\frac{m_1}{m_2}=1+\frac{s_{23}\cot^2\theta_{23}}
{c_{12}s_{12}c^3_{23}}s_{13}\cos\delta.
\end{eqnarray}
Since, the mass ratio $\frac{m_1}{m_2}$ is always smaller than
one, we find that $\cos\delta$ should be negative for $B_1$ and
$B_4$ and positive for $B_2$ and $B_3$, a result that can be
gainfully used to distinguish between mass matrices of class B.
Hence, $B_1$ and $B_4$ (or $B_2$ and $B_3$) will have identical
predictions for $\delta$. Similarly, it can be seen that $B_1$ and
$B_3$ (or $B_2$ and $B_4$) will have, almost, identical
predictions for $\theta_{23}$ if it is near maximality. However,
$B_1$ and $B_4$ (or $B_2$ and $B_3$) can be distinguished from
each other by their predictions for the octant of $\theta_{23}$
and $B_1$ and $B_3$ (or $B_2$ and $B_4$) can be distinguished from
each other by their predictions for the quadrant of $\delta$.
Thus, the four-fold degeneracy in class B is, now, lifted.
Moreover, it follows from Eqs. (26) that the second term should be
suppressed by $s_{13}\cos\delta$ if neutrino mass spectrum is
quasi-degenerate ($m_1 \sim m_2$). For large $\theta_{13}$, the
Dirac phase $\delta$ should be peaked around $90^0$ or $270^0$ to
keep the term $s_{13}\cos\delta$ small. However, for small
$\theta_{13}$, no additional constraint on $\delta$ is obtained.
The fact that $\cos\delta$ should be negative for $B_1$ and $B_4$
type mass matrices and positive for $B_2$ and $B_3$ type mass
matrices for the correct sign of $\Delta m^2_{12}$ results in the
elimination of two quadrants of $\delta$ for each of these
subgroups irrespective of the values of the oscillation parameters
and their errors. In fact, the earlier analyses \cite{2,3} do not
constrain the Dirac phase $\delta$ at all for these subgroup of
mass matrices. In the analysis by Xing \cite{2}, the mass ratios
$\frac{m_1}{m_3}$ and $\frac{m_2}{m_3}$ were examined and this
important constraint on $\delta$ coming from the mass ratio
$\left(\frac{m_1}{m_2}\right)$ was comletely missed as a result.
Another detailed numerical analysis by Guo and Xing \cite{3} based
upon $R_{\nu}$ defined in Eq. (25) does not make use of the solar
mass hierarchy as a result of which no constraints on $\delta$
were obtained for neutrino mass matrices of class B.

\begin{table}[tb]
\begin{center}
\begin{tabular}{|c|ccc|}
\hline
    C.L.    &      $\alpha$      & $M_{ee}$ &       $J$        \\
\hline
 1 $\sigma$ &  $-2.0^0$-$2.0^0$  & $\ge 0.064eV$ & $-0.025$-$0.025$ \\
 2 $\sigma$ &  $-7.7^0$-$7.7^0$  & $\ge 0.037eV$ & $-0.036$-$0.036$ \\
 3 $\sigma$ & $-22^0$-$22^0$ & $\ge 0.023eV$ &  $-0.045$-$0.045$  \\
\hline
\end{tabular}
\end{center}
\caption{The predictions for neutrino mass matrices of class B.}
\end{table}

Now, we present the numerical results of our analysis for mass
matrices of class B. The allowed range of $\delta$ is ($90^0$ -
$270^0$) for types $B_1$ and $B_4$ and ($-90^0$ - $90^0$) for
types $B_2$ and $B_3$. In other words, $\cos \delta$ is negative
for types $B_1$ and $B_4$ and positive for types $B_2$ and $B_3$,
as expected. This important result is valid irrespective of the
ranges of the neutrino oscillation parameters entering in the
analysis and is a consequence of the fact that the sign of $\Delta
m^2_{12}$ is positive. If $\theta_{23}<45^0$, $B_1$ and $B_3$ give
normal hierarchy while $B_2$ and $B_4$ give inverted hierarchy.
Similarly, if $\theta_{23}>45^0$, then  $B_1$ and $B_3$ give
inverted hierarchy while $B_2$ and $B_4$ give normal hierarchy.
Again, this important result is true irrespective of the values of
the neutrino oscillation parameters (and their errors) and follows
from the dependence of mass ratio $\left(\frac{m_1}{m_3} \right)$
on $\tan^2\theta_{23}$ [Cf. Table 2]. As noted earlier, the four
types of neutrino mass matrices within class B differ in their
predictions for $\delta$ and $\theta_{23}$. There is no lower
bound on $\theta_{13}$ in class B. The Majorana-type CP-violating
phase $\beta$, also, remains unconstrained. The predictions for
$\alpha$, $M_{ee}$ and $J$ for class B mass matrices have been
summarized in Table 5. These quantities are the same for mass
matrices of types $B_1$, $B_2$, $B_3$ and $B_4$. It can, also, be
seen that $\alpha$ is restricted to a very small range around zero
and definite lower bounds on $M_{ee}$ are obtained. It is
important to note that $\theta _{23}=45^0$ is disallowed for mass
matrices of class B.

In Fig. 3, the correlation plots for all eight possible cases of B
type mass matrices ($B_{1}$, $B_{2}$, $B_{3}$ and $B_{4}$ with
normal as well as inverted hierarchy) have been shown for
$\alpha$, $\beta $, $\delta$, $\theta _{13}$, $m_{1}$, $M_{ee}$
and $J$. These plots show that B type mass matrices exhibit
eightfold degeneracy in the sense that eight different types of
mass matrices have identical predictions for these parameters.
Figures 3(a-d) depict the correlations of $\alpha$, $\beta$, $J$
and $M_{ee}$ with $\theta_{13}$. It can be seen from Fig. 3(a)
that an upper bound on $\theta_{13}$ constrains $\alpha$ to a very
narrow range. Even at three standard deviations, the allowed range
of $\alpha$ is $-22^{0}\leq \alpha \leq 22^{0}$ [Cf. Table 5].
However, no constraints are obtained on the Majorana phase $\beta$
[Fig. 3(b)]. The correlation plot of $J$ with $\theta_{13}$ shows
a linear behaviour which is a consequence of $\theta _{13}$ being
small and $\delta$ being near $90^{0}$ or $270^{0}$ for most of
the allowed neutrino parameter space.  Since, $M_{ee} \simeq m_1$,
we give the correlation plots for $M_{ee}$ instead of $m_1$ in
Fig. 3(d) where it can be seen that the absolute neutrino mass
scale and 1-3 mixing angles are anti-correlated with each other.
Next, we depict the correlations of the quantities $M_{ee}$ and
$J$ with the Majorana phases $\alpha$ and $\beta$ in figures
3(e-h). The effective Majorana mass $M_{ee}$ diverges at
$\alpha=0$ and $\beta=0$ and so these points are not allowed. A
lower bound on $M_{ee}$ is required to constrain the Majorana
phase $\beta$ which, otherwise, remains unconstrained for mass
matrices of class B [Fig. 3(f)]. It can be seen from Fig. 3(g)
that $J$ and $\alpha$ are directly correlated with each other.
This is due to the fact that $\alpha$ is constrained to a very
narrow range around zero degree so that $\sin\alpha\simeq\alpha$.
The correlation between the two Majorana phases $\alpha$ and
$\beta$ is depicted in Fig. 3(i) which shows that $\beta$ will be
nearly zero for large $\alpha$ and vice-versa. Fig. 3(j) depicts
the correlation between $J$ and $M_{ee}$. One can see that maximal
CP-violation is possible near the lowest possible value of
$M_{ee}$ in class B. It may be worthwhile to mention that the
earlier analysis \cite{3} fails to obtain any constraints on the
ranges of CP-violating phases ($\alpha$, $\beta$ and $\delta$) for
the reasons discussed earlier.

Fig. 4 depicts the correlations which lift the degeneracy between
the neutrino mass matrices of types $B_{1}$ ($B_{4}$) and $B_{2}$
($B_{3}$). Since, the range of Dirac-type CP-violating phase,
$\delta$, is different for $B_{1}$ ($B_{4}$) and $B_{2}$
($B_{3}$), the eightfold degeneracy in B type mass matrices has,
now, been reduced to four-fold. The left panel depicts the
correlation plots for both $B_{1}$ and $B_{4}$ and the right panel
depicts the correlation plots for $B_{2}$ and $B_{3}$. It can be
seen from Fig. 4(a,b) that the range of Dirac-type CP-violating
phase, $\delta$, is $ 90^{0}\leq \delta \leq 270^{0}$
($-90^{0}\leq \delta \leq 90^{0}$) for $B_{1}$ and $B_{4}$
($B_{2}$ and $B_{3}$) type mass matrices, as expected. Also, there
is a strong correlation between Majorana-type CP violating phase,
$\beta$, and Dirac-type CP-violating phase, $\delta$ reinforcing
the zeroth order result $\left(
\beta+\delta=\left(n+\frac{1}{2}\right)\pi \right)$ given in Table
2. It can be seen from Fig. 4(c,d) that $\theta _{13}$ remains
unconstrained in all cases because full range of $\theta _{13}$ is
allowed at $\delta =90^{0}$ and $270^{0}$. However, as we deviate
from these values of $\delta$, $\theta _{13}$ decreases rapidly to
very small values. So, if the CP violation is found to be
non-maximal, $\theta _{13}$ will be constrained to very small
values. However, if CP violation is found to be nearly maximal,
$\theta _{13}$ can be large. Thus, the suppression factor
$s_{13}\cos\delta$ in the first order correction term in the
Taylor expansion for $\left(\frac{m_1}{m_2}\right)$ is small as
expected from the zeroth order approximation for this class.

Fig. 5 depicts the correlations of $M_{ee}$ and $\alpha$ with
$\theta_{23}$. The correlation plots with $\theta _{23}$ can be
used to further reduce the remaining four-fold degeneracy in B
type mass matrices to a two-fold degeneracy which will be lifted
by the determination of the neutrino mass hierarchy. The left
panel (right panel) depicts the correlation plots for $B_{1}(NH)$
and $B_{2}(IH)$ ($B_{1}(IH)$ and $ B_{2}(NH)$). The correlation
plots of $M_{ee}$ and $\alpha$ with $\theta_{23}$ for $B_{3}(NH)$
and $B_{4}(IH)$ (or $B_{3}(IH)$ and $B_{4}(NH)$) will be, almost,
identical to the plots given in Fig. 5 because $B_1$ and $B_3$ (or
$B_2$ and $B_4$) have, almost, identical predictions for
$\theta_{23}$ and, therefore, are not given here. It can be seen
from Fig. 5(a,b) that the effective Majorana mass $M_{ee}$ is
strongly correlated with $\theta_{23}$. Moreover, maximal value of
$\theta _{23}$ is not allowed by the current neutrino oscillation
data since $M_{ee}$ diverges at $\theta_{23}=\frac{\pi}{4}$.
Larger the deviation of $\theta _{23}$ from maximality, smaller is
the value of $M_{ee}$. Also, a large value of $\alpha$ implies
large deviations of $\theta _{23}$ from maximality. The extent of
the deviation of 2-3 mixing from maximality is governed by the
magnitude of the upper bound on $M_{ee}$ while the direction of
the deviation from maximality is governed by the quadrant of
$\delta$ and the neutrino mass hierarchy.

\subsection{Class C}
It can be seen from Table 2 that the mass ratio $\frac{m_1}{m_2}$
is greater than one at zeroth order. However, the higher order
terms in the Taylor series expansion of the ratio
$\frac{m_1}{m_2}$ may make it smaller than one. Detailed numerical
analysis shows that this actually happens and the mass matrices of
class C are marginally allowed. The Taylor series expansion for
$\frac{m_1}{m_2}$ to first order in $s_{13}$ is given by
\begin{equation}
\frac{m_1}{m_2}=\frac{1}{\tan^2\theta_{12}}\left( 1-\frac{\tan
2\theta_{23}}{s_{12}c_{12}}s_{13}\cos\delta \right).
\end{equation}
For $\frac{m_1}{m_2}$ to be smaller than one, the term $(\tan
2\theta_{23} \cos\delta)$ should be positive. Therefore, when
$\theta_{23}<45^0$, we must have $-90^0<\delta<90^0$. Similarly,
when $\theta_{23}>45^0$, we must have $90^0<\delta<270^0$. Just
like class B, the correlation between the octant of $\theta_{23}$
and the quadrant of $\delta$ obtained here for class C is
independent of the ranges of the neutrino oscillation parameters
and is a generic feature of two texture zero mass matrices. The
points $\delta=90^0$, $270^0$ and $\theta_{13}=0^0$ are not
allowed because the first order correction term vanishes at these
points. The mixing angle $\theta_{23}$ should approach $45^0$ as
$\delta$ approaches $90^0$ or $270^0$ and $\theta_{13}$ approaches
$0^0$ so that the term $(s_{13}\tan 2\theta_{23} \cos\delta)$ can
make the ratio $\frac{m_1}{m_2}$ smaller than one. The mass ratio
$\frac{m_1}{m_3}$ is greater than one at the zeroth order.
Therefore, neutrino mass matrices of type C may be expected to be
consistent with inverted hierarchy only. However, we shall see in
the detailed numerical analysis that the normal hierarchy is,
also, allowed marginally.

\begin{table}[tb]
\begin{center}
\begin{tabular}{|c|ccc|}
\hline
            &    $\beta$     &       $J$        &   $M_{ee}$   \\
\hline
 1 $\sigma$ & $-14^0$-$14^0$ & $-0.024$-$0.024$ & $\geq 0.016$ eV \\
 2 $\sigma$ & $-16^0$-$16^0$ & $-0.034$-$0.034$ & $\geq 0.013$ eV \\
 3 $\sigma$ & $-18^0$-$18^0$ & $-0.045$-$0.045$ & $\geq 0.009$ eV \\
\hline
\end{tabular}
\end{center}
\caption{The predictions for $\beta$, $J$ and $M_{ee}$ for
neutrino mass matrices of class C.}
\end{table}

For mass matrices of type C, with inverted hierarchy, the allowed
ranges for the parameters $\beta$, $J$ and $M_{ee}$ are given in
Table 6. All other parameters remain unconstrained. The
correlation plots between various parameters have been depicted in
Figs. 6-8. In Fig. 6, we have given the correlations of $\alpha$,
$\beta$, $\theta_{13}$, $\theta_{23}$, $M_{ee}$ and $J$ with
$\delta$. As discussed earlier, the values $\delta =90^{0}$ and
$270^{0}$ are not allowed [Fig. 6(a)]. Therefore, maximal
CP-violation is not allowed for mass matrices of class C. For the
maximal values of $\delta$, $\alpha$ should be equal to zero.
Therefore, the point $\alpha=0^0$ is, also, not allowed. However,
all other values of $\alpha$ are allowed. It can be seen from Fig.
6(b) that the Majorana phase $\beta$ is a periodic function of
$\delta$ with a period of $180^0$ and $\beta$ can be zero at
$\delta=0^0$ or $180^0$. However, the points ($\beta=0^0$,
$\delta=90^0$) and ($\beta=0^0$, $\delta=270^0$) are disallowed.
The correlation between $\delta$ and $\theta_{13}$ has been
depicted in Fig. 6(c). The correlation between $\delta$ and
$\theta_{23}$ has been depicted in Fig. 6(d) where it can be seen
that the maximal 2-3 mixing is not allowed. However, the mixing
angle $\theta_{23}$ will be below maximality if $\cos \delta$ is
positive ($-90^0$-$90^0$) and above maximality if $\cos \delta$ is
negative ($90^0$-$270^0$). These conclusions are, completely, in
agreement with the results obtained from the Taylor series
expansion. Fig. 6(e) depicts the correlation between $M_{ee}$ and
$\delta$ which yields a lower bound of about 0.02 eV on $M_{ee}$
for $\delta=0^0$ or $180^0$ for neutrino mass matrices of class C.
As $\delta$ deviates from the values $0^0$ and $180^0$, $M_{ee}$
increases and diverges as $\delta$ becomes maximal. The
correlation between $J$ and $\delta$ has been shown in Fig. 6(f).

In Fig. 7, we have shown the correlations of $\alpha$, $\beta$,
$\theta_{23}$ and $J$ with $\theta_{13}$. In Fig. 7(a),
($\theta_{13}$,$\alpha$) correlation has been depicted. Full range
($-90^0$-$90^0$) for $\alpha$ is allowed for $\theta_{13}=0$.
However, it can be seen from Fig. 7(a) that a lower bound on
$\theta_{13}$ of about $4^0$ will constrain $\alpha$ to the range
($-65^0$-$65^0$). As noted earlier, $\beta$ is, already,
constrained to a small range around zero. However, it can be seen
from Fig. 7(b) that $\beta$ can be further constrained if
$\theta_{13}$ is found to be larger than $3^0$. A lower bound on
$\theta_{13}$ larger than $3^0$ will disallow a finite range of
$\beta$ around zero. From the ($\theta_{13}$, $\theta_{23}$)
correlation plot [Fig. 7(c)], it can be seen that larger the value
of $\theta_{13}$, larger will be the deviation of $\theta_{23}$
from maximality. The point ($\theta_{13}=0^0$, $\theta_{23}=45^0$)
is not allowed. The disallowed range of $\theta_{23}$ around
$45^0$ increases with increase in $\theta_{13}$. For example, the
disallowed range of $\theta_{23}$ at $\theta_{13}=6^0$ is
approximately $44^0-46^0$. For small values of $\theta_{13}$, $J$
is proportional to $\theta_{13}$ and the constant of
proportionality is determined by $\sin\delta$ [Cf. Eq. (23) ].
These features can be seen graphically in the correlation plot
between $J$ and $\theta_{13}$ depicted in Fig. 7(d).

Some other important correlation plots have been shown in Fig. 8.
It can be seen from the ($\alpha$, $\beta$) correlation plot [Fig.
8(a)] that the two Majorana phases can not vanish simultaneously.
The correlation between $M_{ee}$ and $\alpha$ has been shown in
Fig. 8(c) where it can be seen that an upper bound on $M_{ee}$
will reject a considerable range of $\alpha$. The ($M_{ee}$,
$\beta$) correlation plot has been given in Fig. 8(d). It can be
seen that the allowed range of $\beta$ (which is $-15^0$-$15^0$
for small values of $M_{ee}$) can be further constrained if
$M_{ee}$ is found to be larger than 0.03 eV. Fig. 8(e) depicts the
correlation between $M_{ee}$ and $\theta_{23}$. It can be seen
that $M_{ee}$ diverges for maximal 2-3 mixing. A lower bound on
$M_{ee}$ can be used to constrain $\theta_{23}$.

For $C$ type mass matrices, with normal hierarchy Dirac-type
CP-violation phase, $\delta$, can have only two values
$\delta=90^{0}$ or $180^{0}$. Majorana-type CP-violating phases
($\alpha$ and $\beta$) and $\theta_{13}$ remain unconstrained.
There is a strong correlation between $\theta_{13}$ and
$\theta_{23}$ in this case. Larger the allowed range of
$\theta_{13}$, larger will be the deviation of $\theta_{23}$ from
maximality. The neutrino mass matrices of class C were examined
analytically in \cite{10} where some of the results reported here
for class C have been obtained.

\subsection{Eightfold degeneracy}

The forthcoming neutrino experiments will aim at measuring the
neutrino mass hierarchy, Dirac type CP violating phase $\delta$
and the deviations of the atmospheric mixing angle from maximality
\cite{19}. The results of these experiments will fall in one of
the following eight categories of phenomenological interest:
\begin{enumerate}
\item NH, $\theta_{23}<45^0$ and $90^0<\delta<270^0$,
\item NH, $\theta_{23}>45^0$ and $-90^0<\delta<90^0$,
\item NH, $\theta_{23}<45^0$ and $-90^0<\delta<90^0$,
\item NH, $\theta_{23}>45^0$ and $90^0<\delta<270^0$,
\item IH, $\theta_{23}>45^0$ and $90^0<\delta<270^0$,
\item IH, $\theta_{23}<45^0$ and $-90^0<\delta<90^0$,
\item IH, $\theta_{23}>45^0$ and $-90^0<\delta<90^0$,
\item IH, $\theta_{23}<45^0$ and $90^0<\delta<270^0$.
\end{enumerate}
The eight cases given above are degenerate for the present
neutrino oscillation data because of the octant degeneracy of
$\theta_{23}$ (i.e. if $\theta_{23}<45^0$ or $\theta_{23}>45^0$),
the intrinsic degeneracy in the sign of $\cos\delta$ and the
two-fold degeneracy in the sign of $\Delta m^2_{23}$. This
degeneracy known as the eightfold degeneracy in the neutrino
parameter space has been studied extensively in the literature
\cite{19,20,21,22}. A specific project named
Tokai-to-Kamioka-Korea (T2KK) two detector complex which will
receive neutrino superbeams from J-PARC has been proposed
\cite{22} to resolve these degeneracies. The intrinsic degeneracy
in the sign of $\cos\delta$ will be resolved by the spectrum
information at T2KK and the degeneracy in the sign of $\Delta
m^2_{23}$ will be resolved by observing the difference in the
earth matter effects between the intermediate and far detectors.
The $\theta_{23}$ octant degeneracy will be resolved by observing
the difference in $\Delta m^2_{12}$ oscillation effects between
the intermediate and far detectors at T2KK. In addition, neutrino
factories \cite{23} and beta-beam experiments \cite{24} have the
potential to resolve the eightfold degeneracy.

\section{Conclusions}

\begin{table}[tb]
\begin{center}
\begin{tabular}{|c|ccc|ccc|}
\hline
 Categories  & &Degeneracies& & &Matrices& \\
\hline
   &  sign of $\Delta m^2_{23}$   &  sign of $\cos2\theta_{23}$ & sign of $\cos\delta$& $A$ &  $B$  & $C$ \\
\hline
 1 & $+$ & $+$ & $-$ &  $A_2$   & $B_1 (NH)$ & $\times$ \\
 2 & $+$ & $-$ & $+$ &  $A_1$   & $B_2 (NH)$ & $\times$ \\
 3 & $+$ & $+$ & $+$ & $\times$ & $B_3 (NH)$ & $\times$ \\
 4 & $+$ & $-$ & $-$ & $\times$ & $B_4 (NH)$ & $\times$ \\
 5 & $-$ & $-$ & $-$ & $\times$ & $B_1 (IH)$ & $\times$ \\
 6 & $-$ & $+$ & $+$ & $\times$ & $B_2 (IH)$ & $C(IH)$  \\
 7 & $-$ & $-$ & $+$ & $\times$ & $B_3 (IH)$ & $C(IH)$  \\
 8 & $-$ & $+$ & $-$ & $\times$ & $B_4 (IH)$ & $\times$ \\
\hline
\end{tabular}
\end{center}
\caption{The consequences of resolving the eightfold experimental
degeneracies for the neutrino mass matrices with two texture zeros.}
\end{table}

If neutrinoless double beta decay searches give positive results
and $M_{ee}$ is measured experimentally or the atmospheric/reactor
neutrino oscillation experiments confirm inverted hierarchy, the
neutrino mass matrices of class A will be ruled out. A generic
prediction of this class of models is a lower bound on
$\theta_{13}$ [Table 3]. If the forthcoming neutrino oscillation
experiments measure $\theta_{13}$ below $3.3^0$, the neutrino mass
matrices of class A will, again, be ruled out. If $M_{ee}$ is
found to be non-zero, the mass matrices of class B or C may be
allowed. The eight different types of mass matrices in class B can
be distinguished from one another by the future experiential data
by the resolution of the eightfold degeneracy of the neutrino
parameter space. The observation of correlations between various
neutrino parameters like $\theta_{13}$, $\theta_{23}$, $\delta$
and $M_{ee}$ etc. will confirm the neutrino mass matrices with two
texture zeros. Moreover, the CP-violation in lepton number
conserving and non-conserving processes will be correlated in a
definite way and there will be only one independent CP-violating
measure for both types of processes.

Main conclusions of the present study have been summarized in
Table 7 which shows how the resolution of the degeneracies of
neutrino parameter space will resolve the degeneracies of the
neutrino mass matrices. As an illustration for the use of Table 7,
we look at the results obtained for category 1 for which the signs
of $\Delta m^2_{23}$ and $\cos \theta_{23}$ are positive and the
sign of $\cos \delta$ is negative. For this category, the neutrino
mass matrices of type $A_2$ and $B_1$ (for normal hierarchy only)
may explain the neutrino data whereas the mass matrices of class C
are ruled out. Other possibilities summarized in Table 7 may be
understood similarly. It is found that the degeneracies in the
neutrino parameter space are inextricably linked to the
degeneracies in the neutrino mass matrix and by finding the
hierarchy, octant of $\theta_{23}$ and sign of $\cos\delta$, the
future experiments will settle the fate of neutrino mass matrices
with two texture zeros. The resolution of the eightfold degeneracy
in the neutrino parameter space will, also, resolve the eightfold
degeneracy of the neutrino mass matrices of class B as can be seen
from figures 4 and 5. Figures 4 and 5, also, illustrate the
decoupling of degeneracies in the sense that two mass matrices can
be degenerate in their $\delta$ predictions but may have different
predictions for $\theta_{23}$ or vice-versa. A similar decoupling
of degeneracies occur in neutrino oscillation experiments with
matter effects \cite{21} where an experiment may be insensitive to
the sign of $\cos\delta$ but can resolve the degeneracy in the
sign of $\cos2\theta_{23}$ or it can be insensitive to the sign of
$\cos2\theta_{23}$ but can resolve the degeneracy in the sign of
$\cos\delta$.

\section{Acknowledgments}

The research work of S. D. and S. V. is supported by the Board of
Research in Nuclear Sciences (BRNS), Department of Atomic Energy,
Government of India \textit{vide} Grant No. 2004/ 37/ 23/ BRNS/
399. S. K. acknowledges the financial support provided by Council
for Scientific and Industrial Research (CSIR), Government of
India. We would like to thank Manmohan Gupta for critical reading
of the manuscript and helpful suggestions.

\newpage

\begin{figure}
\begin{center}
\epsfig{file=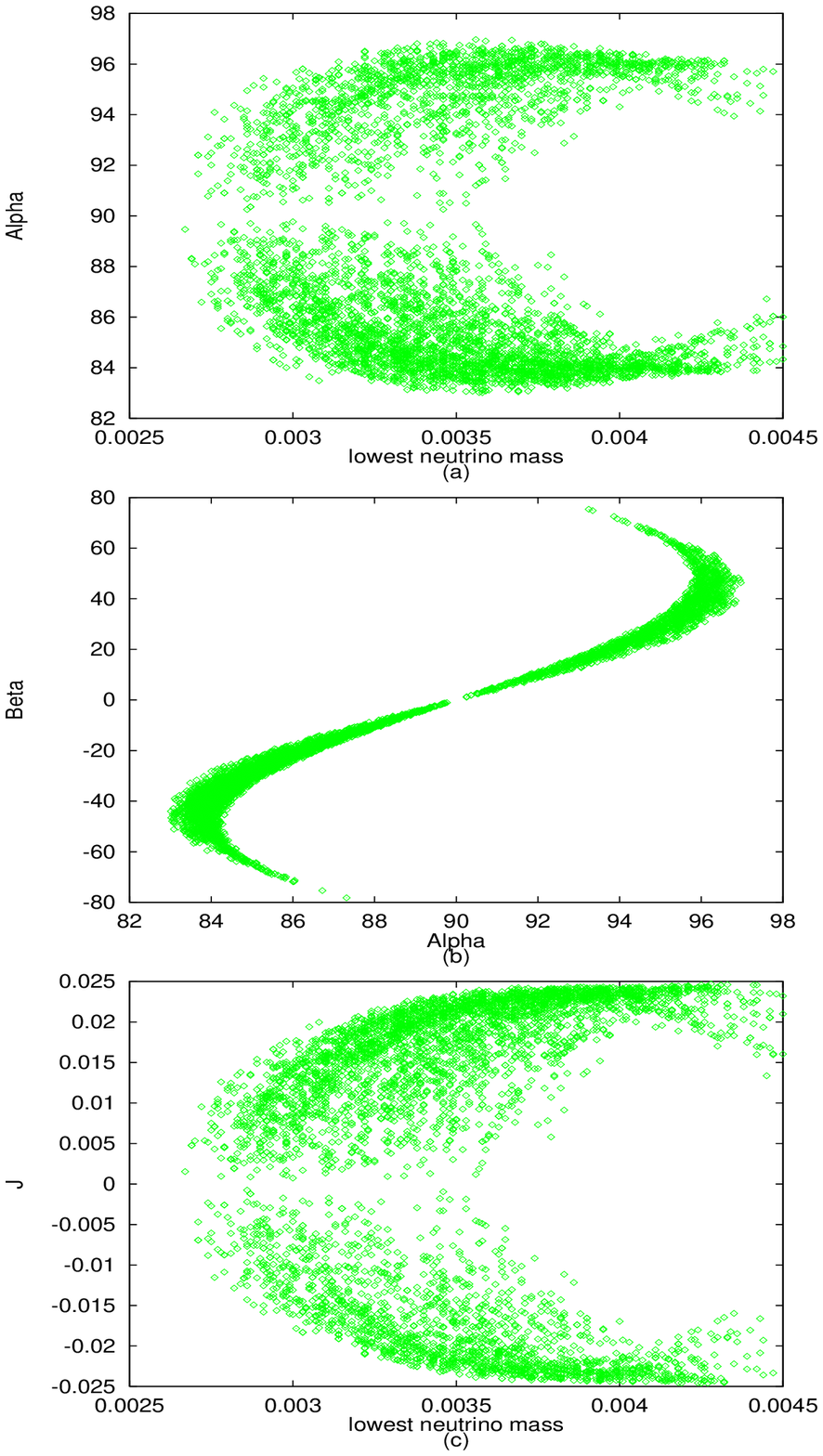,height=20cm,width=15cm}
\end{center}
\caption{ Correlation plots for neutrino mass matrices of class A
at one standard deviation.}

\end{figure}

\begin{figure}
\begin{center}
\epsfig{file=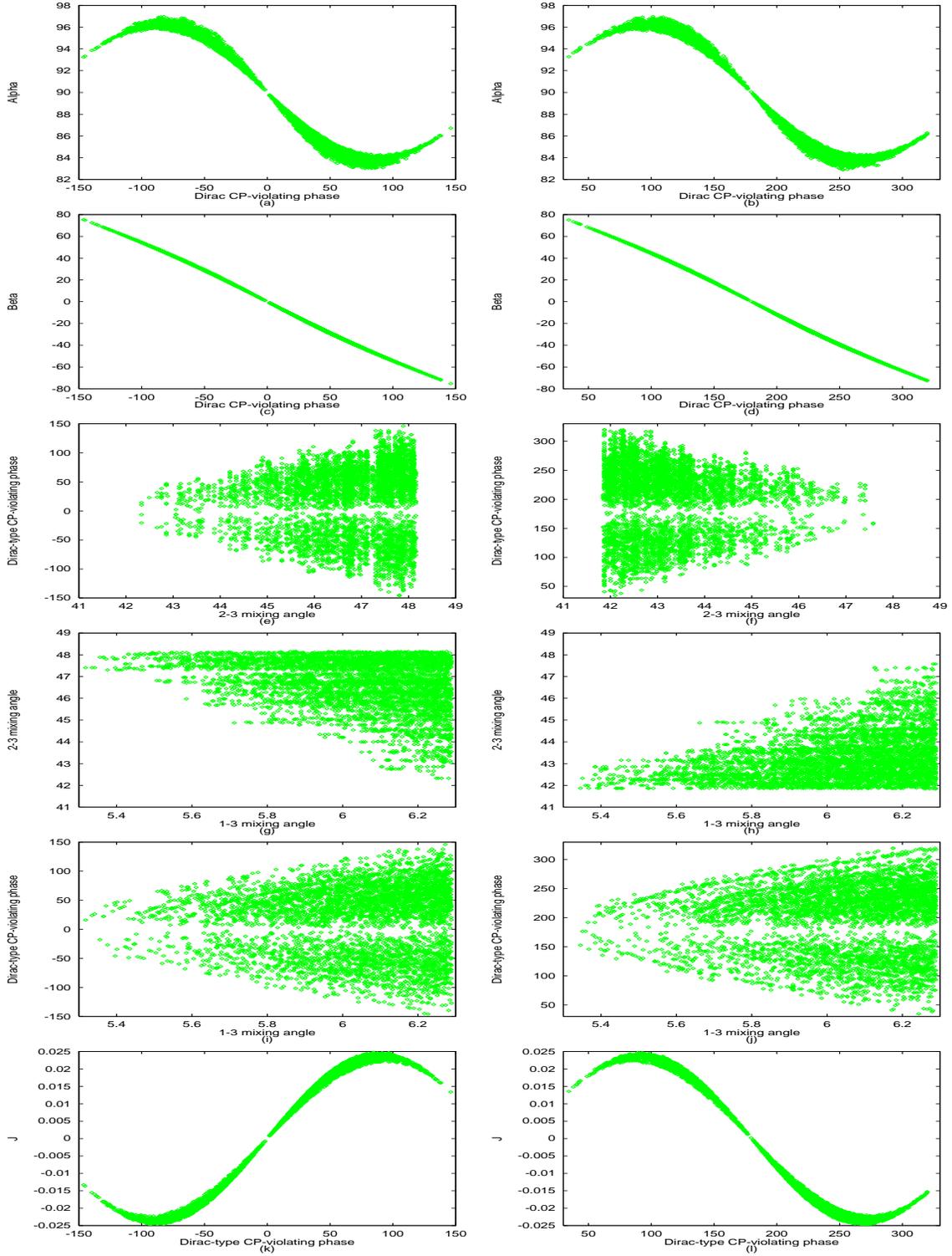,height=20cm,width=15cm}
\end{center}
\caption{ Correlation plots for neutrino mass matrices of type
$A_1$ (left panel) and $A_2$ (right panel).}
\end{figure}

\begin{figure}
\begin{center}
\epsfig{file=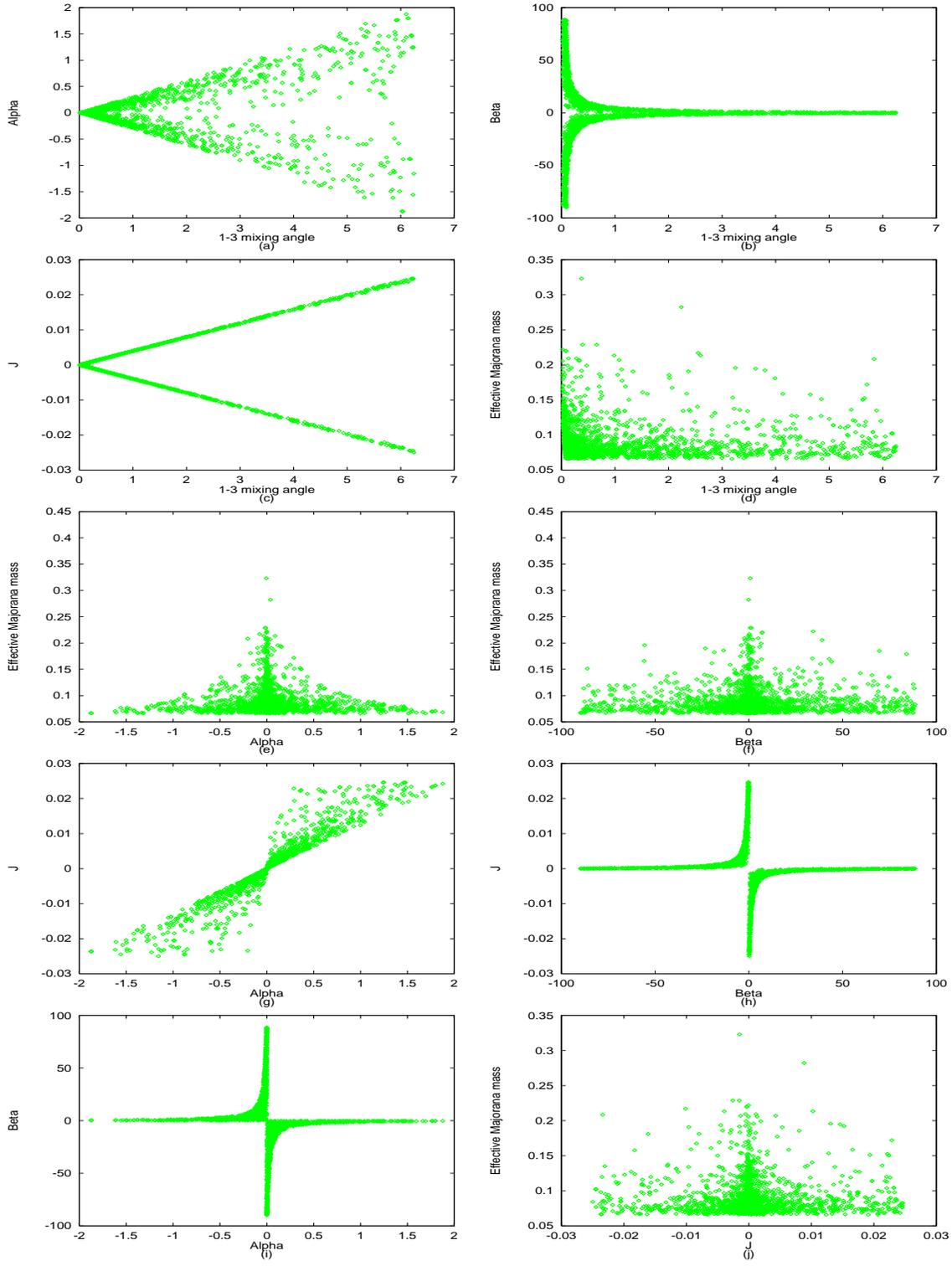,height=20cm,width=15cm}
\end{center}
\caption{ $1\sigma$ correlation plots for $B_{1}$, $B_{2}$, $B_{3}$ and $%
B_{4} $ type mass matrices with normal as well as inverted
hierarchies.}
\end{figure}

\begin{figure}
\begin{center}
\epsfig{file=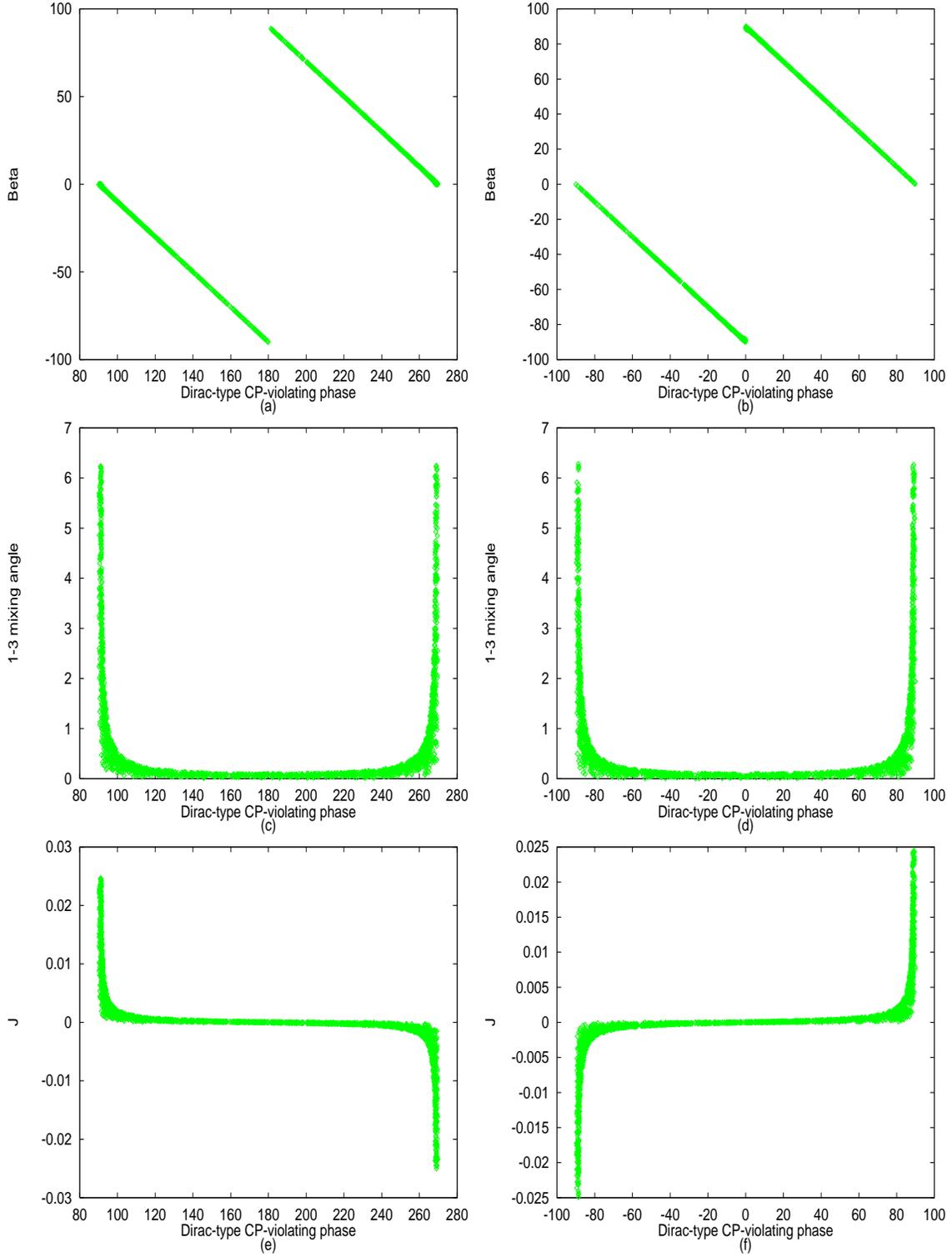,height=20cm,width=15cm}
\end{center}
\caption{ The left panel shows the correlation plots for $B_{1}$
or $B_{4}$ while right panel shows the correlation plots for
$B_{2}$ or $B_{3}$ type mass matrices.}
\end{figure}

\begin{figure}
\begin{center}
\epsfig{file=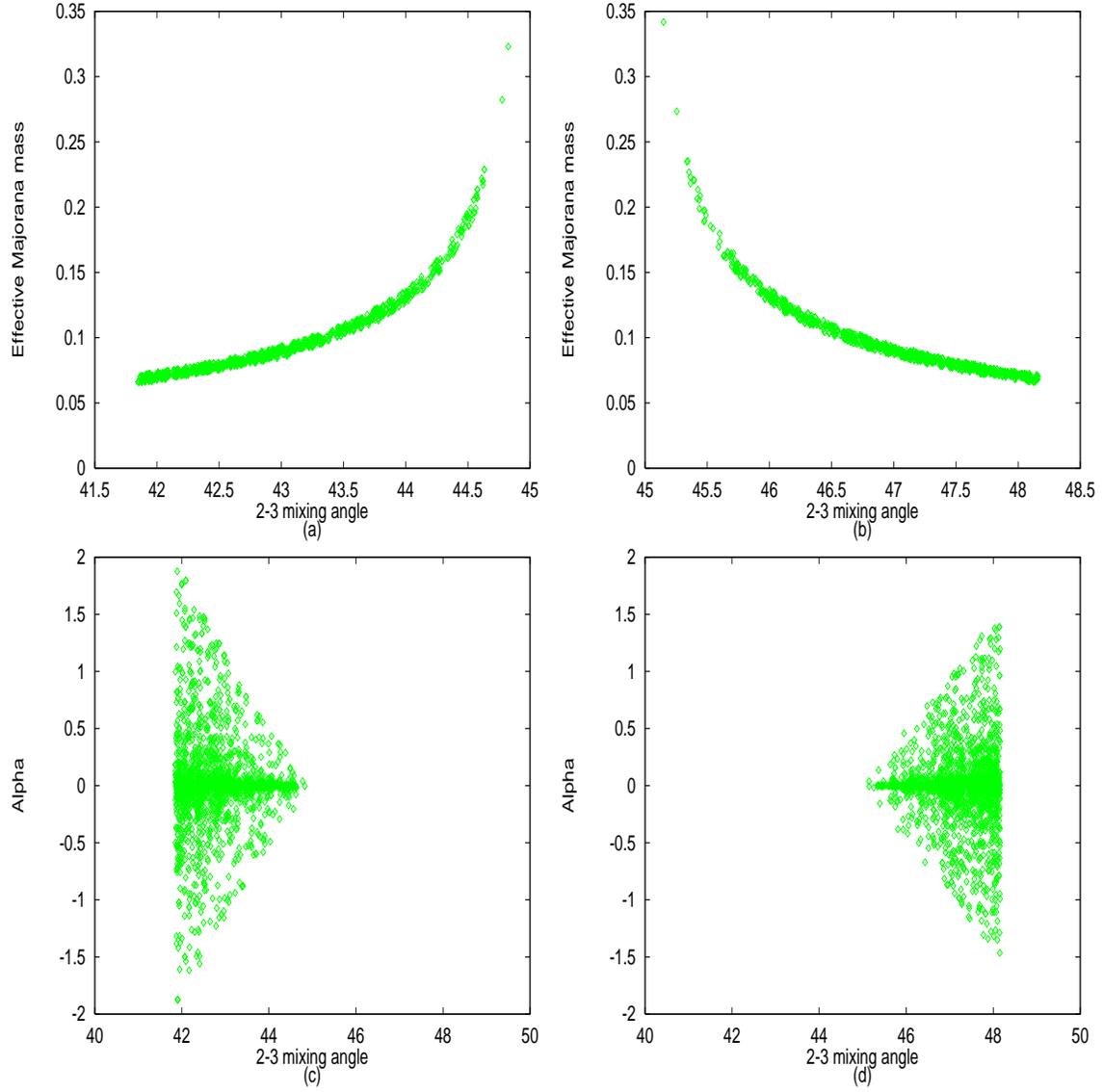,height=15cm,width=15cm}
\end{center}
\caption{ Variation of $M_{ee}$ and $\alpha $ with $\theta_{23}$.
The left [right] panel plots are for $B_{1}(NH)$, $B_{2}(IH)$ [
$B_{3}(IH)$, $B_{4}(NH)$] type mass matrices.}
\end{figure}

\begin{figure}
\begin{center}
\epsfig{file=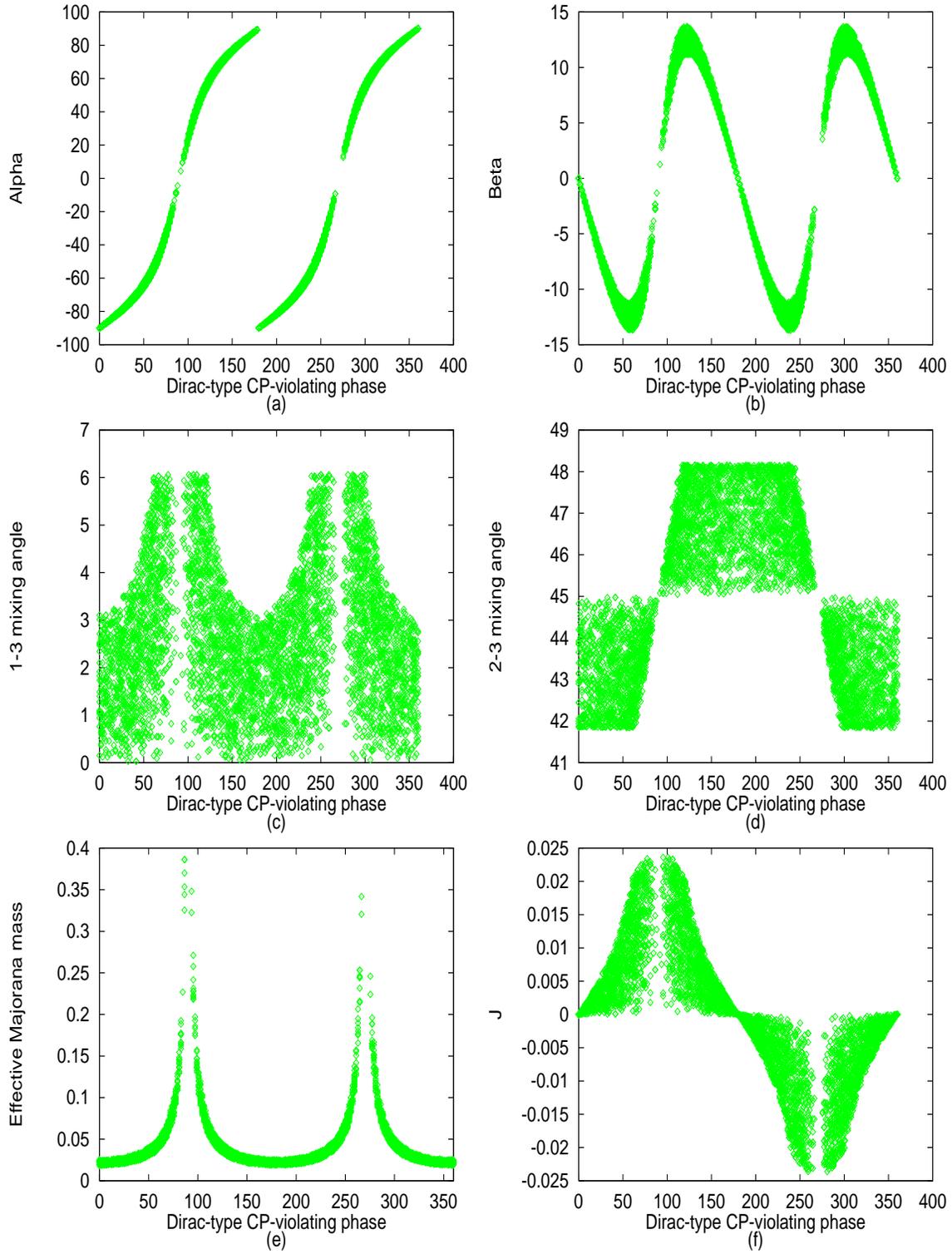,height=20cm,width=15cm}
\end{center}
\caption{The correlation plots for CP-violating phase $\delta$ for
neutrino mass matrices of type C with inverted hierarchy.}
\end{figure}

\begin{figure}
\begin{center}
\epsfig{file=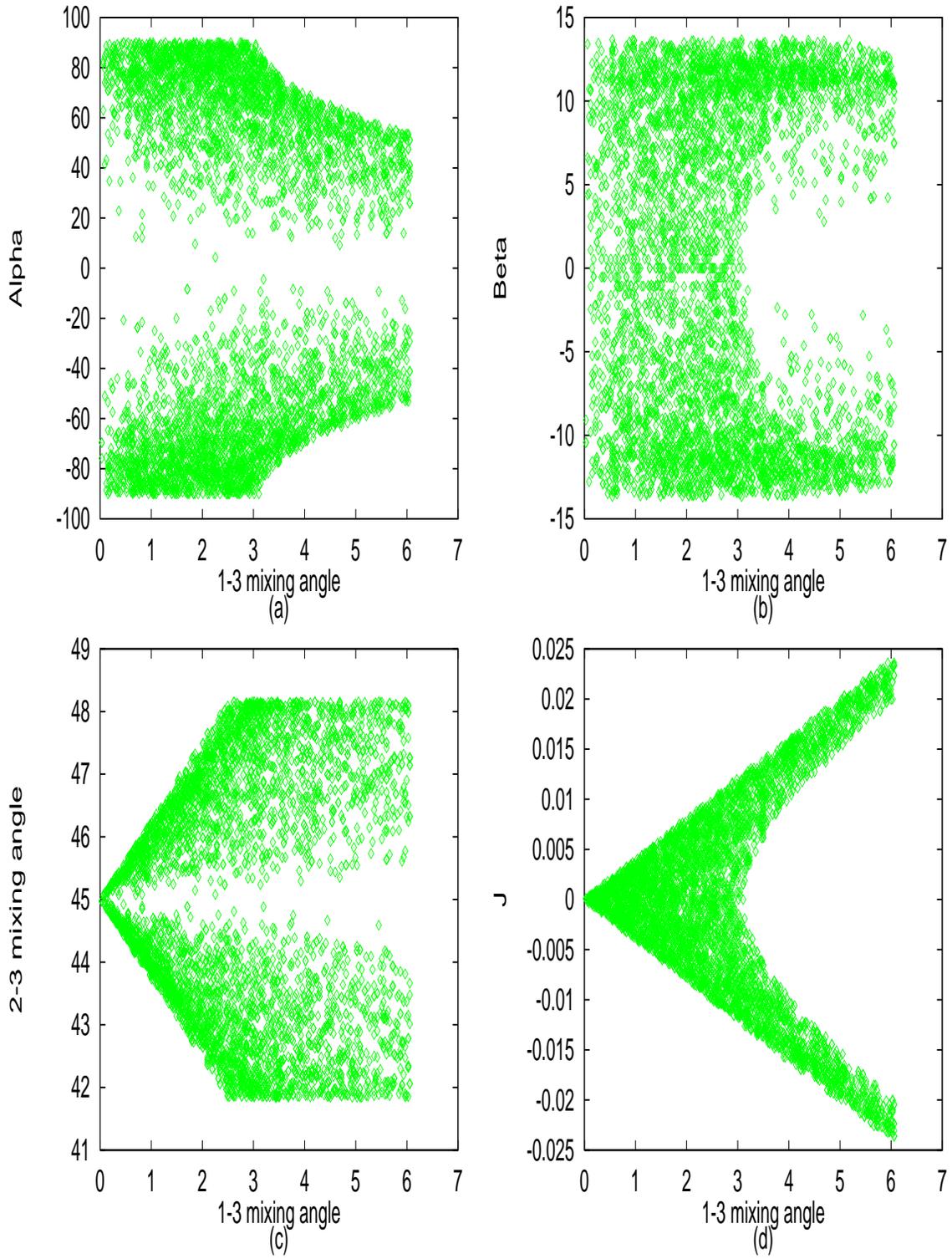,height=20cm,width=15cm}
\end{center}
\caption{The correlation plots for the mixing angle $\theta_{13}$
for neutrino mass matrices of type C with inverted hierarchy.}
\end{figure}

\begin{figure}
\begin{center}
\epsfig{file=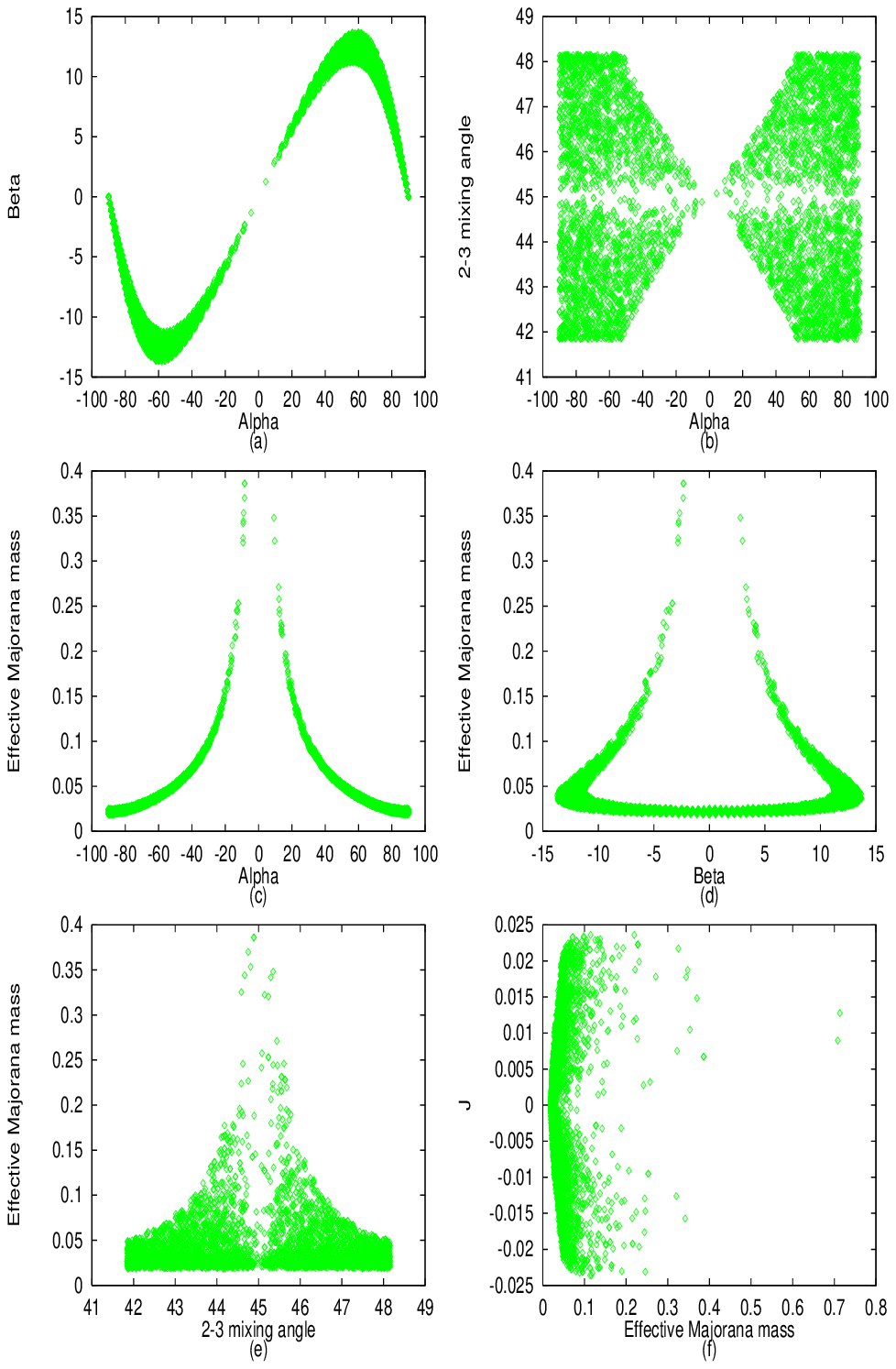,height=20cm,width=15cm}
\end{center}
\caption{Some other important correlation plots for neutrino mass
matrices of type C with inverted hierarchy.}
\end{figure}


\begin{thebibliography}{99}

\bibitem{1} Paul H. Frampton, Sheldon L. Glashow and Danny
Marfatia, \textit{Phys. Lett.} \textbf{B 536}, 79 (2002); Bipin R.
Desai, D. P. Roy and Alexander R. Vaucher, \textit{Mod. Phys.
Lett} \textbf{A 18}, 1355 (2003).

\bibitem{2}  Zhi-zhong Xing, \textit{Phys. Lett.} \textbf{B 530}
159 (2002).

\bibitem{3} Wanlei Guo and Zhi-zhong Xing, \textit{Phys. Rev.} \textbf{D 67},
053002 (2003).

\bibitem{4} Alexander Merle and Werner Rodejohann, \textit{Phys. Rev.}
\textbf{D 73}, 073012 (2006); S. Dev and Sanjeev Kumar,
hep-ph/0607048.

\bibitem{5} S. Dev, Sanjeev Kumar, Surender Verma and Shivani Gupta,
hep-ph/0611313.

\bibitem{6}  G. C. Branco, R. Gonzalez Felipe, F. R. Joaquim and T. Yanagida,
Phys. Lett. \textbf{B 562} 265 (2003); Bhag C. Chauhan, Joao
Pulido and Marco Picariello, \textit{Phys. Rev.} \textbf{D 73},
053003 (2006).

\bibitem{7} Xiao-Gang He and A. Zee, hep-ph/0302201 v2.

\bibitem{8} Suraj. N. Gupta and Subhash Rajpoot, ``Quark Mass Matrices and the Top
Quark Mass'', Wayne State University Preprint, September 1990; S.
N. Gupta and J. M. Johnson, \textit{Phys. Rev.} \textbf{D 44},
2110 (1991); S. Rajpoot, \textit{Mod. Phys. Lett.} \textbf{A 7},
309 (1992); H. Fritzsch and Z.Z. Xing, \textit{Phys. Lett.}
\textbf{B 353}, 114 (1995).

\bibitem{9} P. H. Frampton, M. C. Oh and T. Yoshikawa, \textit{Phys. Rev.}
\textbf{D 66}, 033007 (2002).

\bibitem{10} Walter Grimus and Luis Lavoura, \textit{J. Phys.}
\textbf{G 31}, 693-702 (2005).

\bibitem{11} H. C. Goh, R. N. Mohapatra and
Siew-Phang Ng, \textit{Phys. Rev.} \textbf{D 68}, 115008 (2003).

\bibitem{12} S. Kaneko, M. Katsumata and M. Tanimoto, \textit{JHEP}
\textbf{0307}, 025 (2003).

\bibitem{13}  G. L. Fogli \textit{et al},hep-ph/0506083 v1.

\bibitem{14}M. Maltony, T. Schwetz, M. A. Tortola and J. W. F.
Valle, hep-ph/0405172 v5.

\bibitem{15} J. Nelson, Talk at Neutrino 2006, 13–19 June 2006, Santa Fe, New Mexico,
http://neutrinosantafe06.com.

\bibitem{16} B. Aharmim \textit{et al.} [SNO collaboration], Phys. Rev. C 72, 055502
(2005), nucl-ex/0502021.

\bibitem{17}T. Araki et al. [KamLAND Collaboration], hep-ex/0406035; Talk by
G. Gratta at Neutrino 2004, 14–19 June 2004, Paris,
http://neutrino2004.in2p3.fr.

\bibitem{18} C. Jarlskog \textit{Phys. Rev. Lett.} \textbf{55}, 1039
(1985).

\bibitem{19} E. Abouzaid \textit{et al}, Report of the APS
Neutrino Study Reactor Working Group, 2004.

\bibitem{20} Hisakazu Minakata, Hiroshi Nunokawa, Stephen Parke,
\textit{Phys. Rev.}
\textbf{D 66}, 093012 (2002).

\bibitem{21} Takaaki Kajita, Hisakazu Minakata, Shoei Nakayama, and Hiroshi
Nunokawa,  \textit{Phys. Rev.} \textbf{D 75}, 013006 (2007).

\bibitem{22} Hisakazu Minakata, hep-ph/0701070.

\bibitem{23} Raj Gandhi and Walter Winter, hep-ph/0612158;
A. M. Gago and J. Jones Perez, hep-ph/0611110.

\bibitem{24} E. Fernandez-Martinez, hep-ph/0605101; A. Donini and
E. Fernandez-Martinez, \textit{Phys.Lett.} \textbf{B 641}
432(2006), hep-ph/0603261.


\end{thebibliography}
\end{document}